\documentclass{article}
\usepackage{spconf,amsmath,graphicx,amssymb,delarray,subfigure,verbatim,booktabs}
\usepackage{diagbox, makecell}
\usepackage{multirow}
\usepackage[english]{babel}
\usepackage{lipsum}
\usepackage{setspace}

\title{KS-Net: Multi-band joint speech restoration and enhancement network for 2024 ICASSP SSI Challenge}
\name{\makecell[c]{Guochen Yu$^{\star \dagger}$, Runqiang Han$^{\star \dagger}$\thanks{$^{\dagger}$Contributed equally to this work as co-first authors.}, Chenglin Xu$^{\star}$, Haoran Zhao$^{\star}$, Nan Li$^{\star}$, Chen Zhang$^{\star}$\\
Xiguang Zheng$^{\star}$, Chao Zhou$^{\star}$, Qi Huang$^{\star}$, Bing Yu$^{\star}$}} 

\address{$^{\star}$ Kuaishou Technology, Beijing, China\\
}
\begin{document}
%
\maketitle
\begin{abstract}
\vspace{-0.1cm}
This paper presents the speech restoration and enhancement system created by the 1024K team for the ICASSP 2024 Speech Signal Improvement (SSI) Challenge. Our system consists of a  generative adversarial network (GAN) in complex-domain for speech restoration and a fine-grained multi-band fusion module for speech enhancement. In the blind test set of SSI, the proposed system achieves an overall mean opinion score (MOS) of 3.49 based on ITU-T P.804 and a Word Accuracy Rate (WAcc) of 0.78 for the real-time track, as well as an overall P.804 MOS of 3.43 and a WAcc of 0.78 for the non-real-time track, ranking $1^{st}$ in both tracks.
\end{abstract}
\vspace{-0.2cm}

\vspace{-0.2cm}
\section{Introduction}
\vspace{-0.3cm}
\label{sec:intro}
During audio communication, speech signals recorded by microphones may be severely degraded by various distortions, including environmental noise, reverberation, coloration, discontinuity and loudness. Although existing methods have shown effectiveness in the noise suppression, restoring high-quality speech with the presence of multiple simultaneous distortions still remains challenging. 
The ICASSP SSI Challenge aims to restore the high-qulity speech signal from the captured signal distorted by the aforementioned complex acoustic conditions in a real-time communication system \cite{cutler2023icassp}.
In this paper, we propose a multi-band joint restoration-and-enhancement framework called \textbf{KS-Net} to address the aforementioned distortions. Firstly, the framework adopts a complex-domain based generative adversarial network (GAN) to restore the coarse-grained high-quality speech, aiming to remove various types of distortions. Subsequently, the remaining transient noise and artifacts after the restoration GAN are eliminated by a multi-band fusion network for speech enhancement. The proposed KS-Net approach achieves the $1^{st}$ places in both real-time and non-real-time tracks in the 2024 SSI challenge.
\begin{figure*}[ht!]
    \centering
    \centerline{\includegraphics[width=1.7\columnwidth]{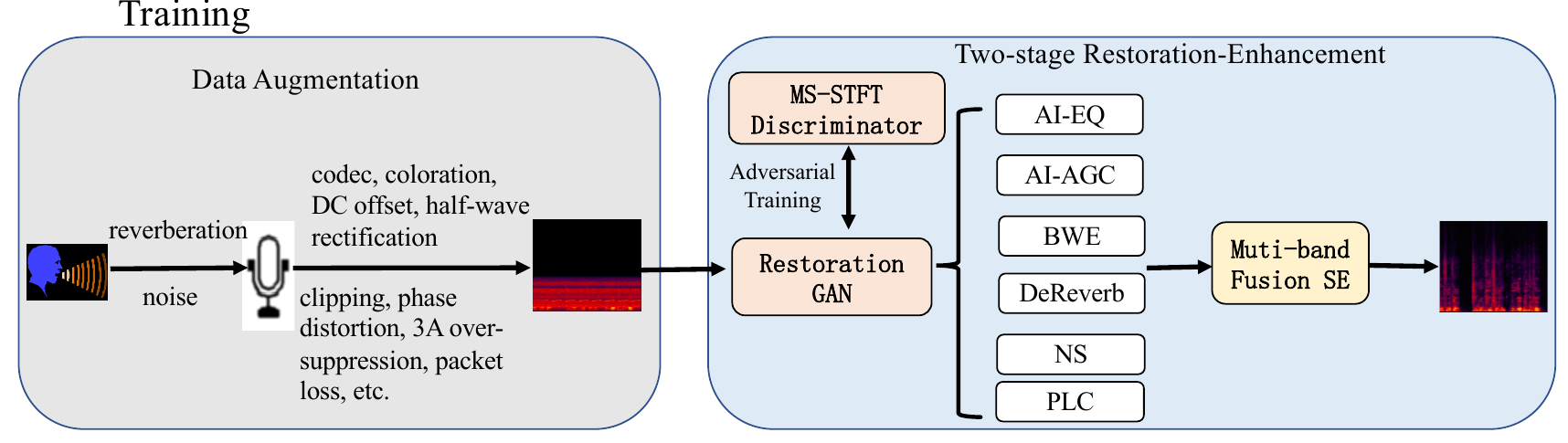}}
    \vspace{-0.2cm}
    \caption{The flowchart of the proposed KS-Net.}
    \vspace{-0.5cm}
    \label{fig:diagram-system}
  
\end{figure*}
\vspace{-0.3cm}
\section{System Architecture}
\vspace{-0.3cm}
\label{sec:format}

As illustrated in Fig~{\ref{fig:diagram-system}}, the training procedure of our system consists of two stages. The first stage involves the primary restoration GAN, which aims to address various distortions, such as band-width extension (BWE), speech distortion restoration, loudness adjustment, packet loss concealment (PLC), noise suppression (NS) and dereverberation (Dereverb). In the second stage, the multi-band fusion SE (MF-Net) focuses specifically on removing residual transient noise and artifacts generated by the restoration GAN.
\vspace{-0.3cm}
\subsection{Complex-domain based restoration GAN}
\vspace{-0.2cm}
\label{sec:pagestyle}
Inspired by GAN-related methods presented in ICASSP 2023 SSI-Challenge{\cite{chen2023gesper, zhu2023ssi}}, we proposed a restoration GAN to restore the coarse-grained high-quality speech in the complex domain. To facilitate information exchange between low frequency bands and high frequency bands, we divide the full-band complex spectrum into three subbands and concatenate them along the channel dimension as inputs to the generator.
The restoration GAN's generator follows a convolutional encoder-decoder architecture. It includes five groups of squeezed temporal convolution modules (S-TCMs) and two residual time-frequency LSTM (TF-LSTM) for sequence modeling in the bottleneck.
The encoder consists of five convolutional layers, with a dilated dense-block inserted after the first convolutional layer, which contains five dilated convolutional layers. On the other hand, the decoder comprises five corresponding transposed convolutional layers and a corresponding dilated dense-block.
To improve temporal modeling, each group of S-TCMs is composed of four S-TCN blocks with exponentially increasing dilation rates. After the last S-TCMs, two TF-LSTM are stacked. The time-LSTM focuses on capturing long-term dependencies along the time axis, while the bi-directional frequency-LSTM targets the long-term dependencies along the frequency axis.


Regarding the discriminators, multi-resolution STFT discriminators are used to capture various spectral patterns in speech signals\cite{kong2020hifi}. The training loss is a combination of a reconstruction loss, an adversarial loss, and a feature match loss. The reconstruction loss includes both a multi-resolution STFT loss and a waveform loss.

\vspace{-0.5cm}
\subsection{Multi-band fusion SE}
\vspace{-0.3cm}

The restoration GAN's output lacks certain spectral details in the low-frequency parts and still has some leftover transient noise. To address this, we introduce the MF-Net, a multi-band fusion enhancement module. This module aims to restore the spectral details and eliminate the remaining noise.

Based on the insights from our prior study \cite{yu2023fsi}, we have developed MF-Net - a two-stage solution that simultaneously restores the overall full-band spectral pattern and the intricate sub-band spectral details. 
In the first stage, we apply coarse denoising to the compressed ERB-scaled spectrum. This step effectively reduces noise and enhances the general features.
In the second stage, due to the varying spectral characteristics among different frequency bands, we employ two distinct sub-networks to refine the low-frequency and high-frequency bands separately. 


\vspace{-0.4cm}
\section{Experiments}
\label{sec:typestyle}

\vspace{-0.3cm}
\subsection{Data Augmentation}
\vspace{-0.2cm}
The training set for this project is constructed using multiple sources: the DNS Challenge dataset\cite{dubeyicassp}, the VCTK corpus\cite{veaux2017cstr} and our own private datasets.  
 Specifically, we create a large dataset of 800 hours of paired data. In this dataset, we add different distortions, such as various types of noise, reverberation, clipping, packet loss, low-pass filtering, traditional noise reduction algorithms, and low bit-rate codecs, to the original clean full-band speech. We add these distortions with different probabilities to simulate real-life scenarios.
During the training stage for MF-Net, transient noises are further added to the outputs of the restoration GAN with signal-to-noise ratios (SNRs) ranging from -5dB to 10dB.

\vspace{-0.4cm}
\subsection{Experimental Setup}
\vspace{-0.2cm}
\label{sec:majhead}
In our experiments, we utilized a Hanning window with a 20 ms window length and a 10 ms frame shift. It is worth noting that the difference between the submitted system for SSI real-time track (\textbf{KS-Net-1}) and non-real-time track (\textbf{KS-Net-2}) is that the generator and MF-Net in KS-Net-2 is designed with a large amount of network parameters and computational complexity. KS-Net-1 has a total of 15.64 million(M) trainable parameters and its real time factor (RTF) is 0.42 on an Intel Core i5 Quadcore CPU clocked at 2.4 GHz, while KS-Net-2 has a total parameter number of 19.15 M and an RTF of 0.62.
\vspace{-0.7cm}
\subsection{Results and analysis}
\vspace{-0.2cm}
Table~{\ref{tbl:results}} presents the results of the P.804 subjective test conducted on the official blind set. It is evident that both KS-Net-1 and KS-Net-2 exhibit considerable improvements across all subjective metrics when compared to the noisy signals. This suggests that our proposed system effectively enhances speech quality by alleviating distortions introduced by various factors such as noise, reverberation, coloration, discontinuity, and loudness. 
Moving forward, our future research endeavors will focus on further exploring the underlying factors contributing to the superior performance of real-time KS-Net-1 in contrast to non-real-time KS-Net-2.



 		\renewcommand\arraystretch{1.2}
		\begin{table}[t!]
			\setlength\tabcolsep{3.0pt}
			\caption{ITU-T P.804 MOS and WAcc results on the SSI 2024 blind test set.}
                \vspace{0.2cm}
			\centering
			\scalebox{0.8}{
				\begin{tabular}{l|ccccccc|l}
      
                    \hline
    			\multirow{2}*{System} &\multicolumn{7}{c|}{\textbf{MOS}}  &{\textbf{WAcc}} \\
                    \cline{2-8}
    			  &COL &DISC&LOUD &NOISE &REVERB   &{SIG}  &ORVL &{(\%)}\\
				\hline

                    Noisy  &3.34 &3.70 &3.78 &3.21 &3.40 &3.05 &2.58 &\textbf{82.68} \\ 
     		    KS-Net-1  &\textbf{4.08} &\textbf{3.89} &\textbf{4.34} &\textbf{4.30} &\textbf{4.27} &\textbf{3.83} &\textbf{3.49} &78.23 \\ 
                    KS-Net-2  &4.01 &\textbf{3.89} &\textbf{4.34} &4.28 &4.24 &3.74 &3.43 &77.30 \\ 
                \hline

				\end{tabular}
			}
			\label{tbl:results}
            \vspace{-0.3cm}
		\end{table}

\vspace{-0.1cm}
\bibliographystyle{IEEEbib}
\vspace{-0.2cm}
\let\oldbibliography\thebibliography 
\renewcommand{\thebibliography}[1]{ 
  \oldbibliography{#1}
  \setlength{\itemsep}{-1pt} 
}
\begin{spacing}{0.9} 
    \bibliography{myrefs} 
\end{spacing} 

\vfill\pagebreak


\end{document}